\documentclass[12pt,a4paper]{article}
\usepackage[latin1]{inputenc}

\usepackage{amsmath, amsfonts, amssymb, graphicx, float}
\DeclareGraphicsExtensions{.pdf,.png,.jpg,.ps,.gif,.tif,.eps}

\usepackage{amsmath}
\usepackage{amsfonts}
\usepackage{amssymb}
\usepackage{multirow}
\usepackage[english]{babel}
\usepackage{lscape}
\usepackage[left=2.5cm,right=2.5cm,top=2cm,bottom=2cm]{geometry}
\usepackage{float}
\usepackage{tabto}

 \usepackage{afterpage}

\usepackage{fancyhdr}
\usepackage{caption}
\author{J.L. Raath, M.S. Potgieter, R.D. Strauss, A. Kopp}
\title{The effect of magnetic field modifications on the modulation of cosmic rays in the heliosphere}
\begin{document}
\maketitle
\section*{Abstract}

A numerical model for the solar modulation of cosmic rays, based on the solution of a set of stochastic differential equations, is used to 
illustrate the effects of modifying the heliospheric magnetic field, particularly in the polar regions of the heliosphere.  To this end, the 
differences in the modulation brought about by each of three choices for the heliospheric magnetic field, i.e. the unmodified Parker field, 
the Smith-Bieber modified field, and the Jokipii-K\'{o}ta modified field, are studied.  It is illustrated that both the Jokipii-K\'{o}ta and Smith-Bieber 
modifications are effective in modifying the Parker field in the polar regions.  In addition, it is argued that the modification of Smith and Bieber 
is based on observational evidence and has a firm physical basis, while these motivations are lacking in the case of the Jokipii-K\'{o}ta modification.  
From a cosmic ray modulation point of view, we found the Smith-Bieber modification to be the most suitable choice for modifying the 
heliospheric magnetic field.  The features and effects of these three modifications are illustrated both qualitatively and quantitatively.  
It is also shown how the Smith-Bieber modified field can be applied in cosmic ray modulation models to reproduce observational cosmic 
ray proton spectra from the PAMELA mission during the solar minimum of 2006 - 2009.  These results are compared with those obtained 
in previous studies of this unusual solar minimum activity period and found to be in good qualitative agreement.
\\ \\
\textbf{Keywords}  Solar modulation $\cdot$ Cosmic rays $\cdot$ Stochastic differential equations $\cdot$ Heliospheric magnetic field $\cdot$ PAMELA $\cdot$ Solar minimum

\section{Introduction}

The intensities of Galactic cosmic rays (CRs) entering the heliosphere
are modulated by a number of physical mechanisms, including
convection, particle drifts and diffusion, as well as adiabatic energy changes.  Modeling these mechanisms and the associated modulation
of CRs is done primarily by employing numerical modulation models, since analytical models are limited to only a few simplified cases.

The numerical modulation model used in this study is based on the solution of an appropriate set of stochastic differential equations (SDEs).  
Studies in the past have mainly employed numerical models based on finite difference approaches, the alternating direction implicit 
(ADI) being the most extensively utilised scheme.  However, after Zhang (1999) successfully applied SDEs to study CR modulation 
in the heliosphere, the use of such models has become popular.  SDE-based models have been used to study the transport of pickup 
ions (Fichtner \textit{et al.}, 1996), solar energetic particles (Dr\"{o}ge \textit{et al.}, 2010), the propagation of Galactic and Jovian electrons 
(Strauss \textit{et al.}, 2011a), and other aspects of CR  modulation (e.g. Pei \textit{et al.} 2010; Strauss \textit{et al.}, 2011b, 2012;  Luo \textit{et al.}, 2011, 2013a, 
2013b).  Some historical and contemporary models are discussed by e.g. Yamada, Yanagita and Yoshida (1998), Alanko-Huotari \textit{et al.} (2007), Bobik \textit{et al.} 
(2012), and Kopp \textit{et al.} (2012) who also discussed the mathematical and practical implementation of SDEs for a variety of general 
transport equations.

The SDE-approach leads to numerical models that are ideal for implementation on multiple processors, allowing for parts of the numerical 
code to be executed simultaneously and independently, saving a great deal of computation time.  Besides the parallel use of multiple central processing units (CPUs)
and computer clusters, high performance calculations on graphics processing units (GPUs) have also become available during the last
years;  in such a GPU-accelerated implementation to study the solar modulation of CRs, presented by Dunzlaff, Strauss and Potgieter (2015), a performance
increase of a factor of about 10 - 60 was found when compared to the CPU-version of the same algorithm.  ADI-based models, on the other hand, 
only allow for a linear execution of the numerical code so that improvement in computation time cannot be achieved by these means.  Furthermore, 
SDE-based models, being independent of a spatial grid, have also displayed remarkable numerical stability.  This makes it possible to incorporate 
detailed descriptions of the heliosphere and heliospheric structures into these models.  Examples of such implementations include the works 
of Strauss \textit{et al.} (2011b), where the relative motion of Jupiter $-$ as the source of Jovian electrons $-$ was accounted for;  Strauss \textit{et al.} (2012), where a fully three-dimensional wavy heliospheric current 
sheet (HCS) was implemented;  Luo \textit{et al.} (2013a), where acceleration of the anomalous component and the re-acceleration of Galactic CRs by 
the solar wind termination shock (TS) were included; and Luo \textit{et al.} (2011, 2013b), where time-varying modulation parameters were implemented.  ADI-based models, on the 
other hand, are notoriously unstable when solving differential equations in higher dimensions and the complexity of these models are thus restricted.
In addition to these considerations, SDE-based models are especially powerful in their ability to visualise the modulation process, being  
able to calculate pseudo-particle trajectories or $-$ equivalently stated $-$ follow the evolution of individual phase space (density) elements; it is important to note 
that these trajectories are not the trajectories of actual particles, but can at most be interpreted to obtain an indication of how the modulation of actual 
particles would proceed.  The ability to calculate these trajectories leads to the possibility of calculating additional modulation features which were 
previously not possible, e.g. the propagation times and energy losses of particles in the heliosphere.  

The heliospheric magnetic field (HMF) plays a central role in the modulation mechanisms pointed out above and, as such, the choice of the HMF profile
is a very important question in modulation studies.  In this paper, we apply the SDE-based model to investigate the effect of different choices for the HMF.  
We consider the unmodified Parker HMF (Parker, 1958) and two modifications to this field, namely that of Jokipii and K\'{o}ta (1989) and that of Smith and 
Bieber (1991).  The focus is on the modification of Smith and Bieber (SBM), and it is argued that this modification should be preferred when modifying Parker 
like HMFs.  To this end the numerical model is applied, employing the SBM, to reproduce the Galactic proton spectra observed at Earth by the PAMELA mission during 
the peculiar solar minimum of 2006 to 2009.  The results are then compared to those of other authors who have studied this minimum.

\section{The transport model for Galactic protons}

An equation including most of the relevant heliospheric modulation mechanisms was derived by Parker (1965) and is known as the 
Parker transport equation (TPE), given by
\begin{equation}\label{eq:TPE}
\dfrac{\partial f}{\partial t} = -\left( \vec{V}_{\mathrm{sw}}+\langle\vec{v}_{\mathrm{d}}\rangle \right)\cdot\nabla f +\nabla\cdot\left( \mathrm{\bf{K}}_{\mathrm{s}} 
\cdot \nabla f \right) +\dfrac{1}{3}\left(\nabla\cdot \vec{V}_{\mathrm{sw}}\right)\dfrac{\partial f}{\partial \ln p}+Q,
\end{equation}
in terms of the omnidirectional distribution function $f\left(\vec{r},p,t\right)$, where $\vec{r}$ is the position, $p$ the particle 
momentum, and $t$ the time.  The first term on the right hand side describes the process of convection via the expanding solar 
wind (SW), with velocity 
$\vec{V}_{\mathrm{sw}}$; 
the second term represents the effects of gradient and curvature drifts, where $\langle\vec{v}_{\mathrm{d}}\rangle$ is 
the pitch-angle averaged guiding centre drift velocity;  the third term describes particle diffusion through the symmetric 
diffusion tensor $\textbf{K}_{\mathrm{s}}$;  the fourth term accounts for adiabatic energy changes; the final term on 
the right hand side, $Q$, includes any source terms, if required.  For a detailed description of these modulation processes, 
see e.g. Potgieter (2013).

\begin{figure}
\begin{center}
\includegraphics[height=5.0 in,width=3.8 in, angle = 90]{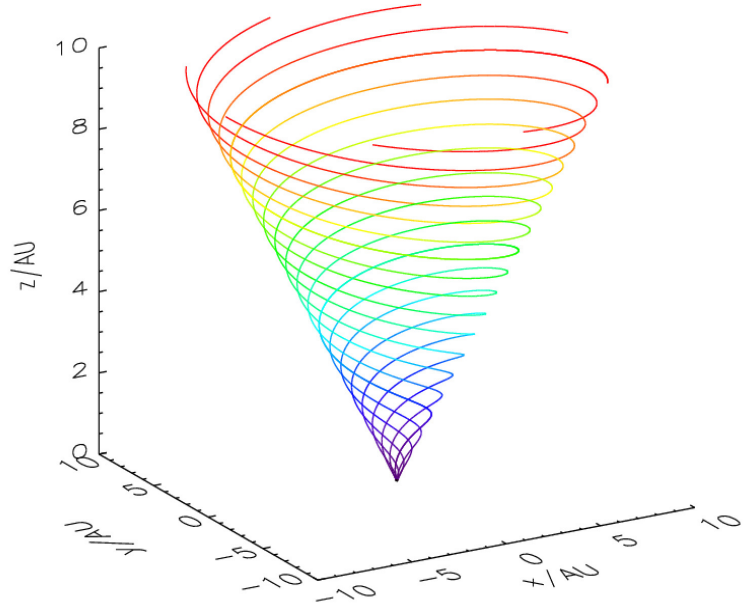}
\caption{\small \sl Magnetic field lines for the case of the Parker HMF, originating from consecutive azimuthal angles and a latitude of $\theta=45^\circ$;  shown in
Cartesian coordinates and over the first 10 AU
from the Sun, which is located at $\left(X,Y,Z\right)=\left(0,0,0\right)$;  the colour scale corresponds to $r^2|B|$.  The spiral structure of this field is clearly depicted.}\label{fig:PHMF}
\end{center}
\end{figure}

The geometry, magnitude and turbulence of the HMF play a key role in most of these modulation processes, in particular in determining 
particle drifts, as will be shown and emphasised below. In most CR modulation models the straight-forward Parker HMF (Parker, 1958) 
is used.  This HMF profile is valid for $r\geq r_\odot$, where $r_\odot=0.005$ AU is the solar radius, and is expressed 
in spherical coordinates $\left(r,\theta,\phi\right)$ as
\begin{equation}
\vec{B}=B_0\left[\dfrac{r_0}{r}\right]^2\left(\textbf{e}_r-\tan\psi\textbf{e}_\phi\right),
\end{equation}
where $\textbf{e}_r$ and $\textbf{e}_\phi$ are unit vectors in the radial and azimuthal directions respectively,
$B_0$ is the magnitude of the HMF at $r_0=1$ AU (i.e. at Earth), and the spiral angle $\psi$ is defined by
\begin{equation}\label{eq:PHMFspiral}
\tan\psi=\dfrac{\Omega\left(r-r_\odot\right)}{V_{\mathrm{sw}}}\sin\theta = \Psi.
\end{equation}
In the above expression, $\Omega=2.66\times10^{-6}$ $\mathrm{rad.s^{-1}}$ is the average angular rotation speed of the Sun.  The 
magnitude of the Parker HMF is then given by
\begin{equation}\label{eq:PHMF}
B=B_0\left[\dfrac{r_0}{r}\right]^2\sqrt{1+\Psi^2}.
\end{equation}
Figure \ref{fig:PHMF} shows the magnetic field lines in the case of the Parker HMF, converted to Cartesian coordinates, and its spiral structure is clearly illustrated over the first 10 AU 
from the Sun, which is located at $\left(X,Y,Z\right)=\left(0,0,0\right)$;  these lines originate at consecutive azimuthal angles and a latitude 
of $\theta=45^\circ$.

The Parker HMF, however, leads to very large gradients in particle intensities $j=fP^2$, with $P$ the particle rigidity, in the polar regions;   this makes for a gross over-estimation 
of drift effects in these regions, as was originally pointed out by Potgieter, le Roux and Burger (1989);  see also Potgieter (2013).  An example of such drift dominated 
modulation is found in Jokipii and Kopriva (1979).  The Parker HMF has therefore been modified in several studies to scale down these effects in the polar 
regions;  this forms the topic of Section \ref{sec:HMFmod}.

The SW profile selected is rather simple, accelerating from zero to a constant and supersonic velocity within the first 0.3 AU from the Sun.  
The only other radial structure occurs at the position of the TS, i.e. at $r=r_{\mathrm{TS}}$, where $V_{\mathrm{sw}}$ drops off to subsonic levels;  The SW also has 
a latitudinal profile, which accounts for a slow SW of $\sim$400 $\mathrm{km.s^{-1}}$ in the equatorial regions and a fast SW of $\sim$800 
$\mathrm{km.s^{-1}}$ in the polar regions, in accordance with Ulysses observations during solar minimum (Phillips \textit{et al.}, 1995).

Particle diffusion is realised through the diffusion tensor $\textbf{K}_{\mathrm{s}}$ in Eq. (\ref{eq:TPE}), of which each element is known as a diffusion coefficient.
An approach similar to that of Potgieter \textit{et al.} (2014) is followed;  this will also allow for a direct comparison with their
modulation modeling, which did not use the SBM.
The parallel diffusion coefficient $\kappa_{||}$, i.e. parallel to the mean HMF, is assumed to have a power-law dependence on particle rigidity,  while its spatial dependence
is assumed to be inversely proportional to the magnitude of the magnetic field, so that
\begin{equation}\label{eq:kpar}
\kappa_{||}=\kappa_{||,0}\beta\dfrac{B_0}{B}\left[ \dfrac{\left(\dfrac{P}{P_0^{'}}\right)^{a_3}+\left(\dfrac{P_{\mathrm{k}}}{P_0^{'}}\right)^{a_3}}{1+\left(\dfrac{P_{\mathrm{k}}}{P_0^{'}}\right)^{a_3}} \right]^{\dfrac{a_2-a_1}{a_3}}\left(\dfrac{P}{P_0^{'}}\right)^{a_1},
\end{equation}
where $\kappa_{||,0}$ is a constant in units of $6\times10^{20}$ $\mathrm{cm^2.s^{-1}}$, with $P_0^{'}=1$ GV and $B_0=1$ nT added to obtain the correct dimensions;  $\beta=v/c$
is the ratio of particle speed $v$ to the speed of light $c$.  
Here $a_1$ and $a_2$ are dimensionless
constants and determine the slope of the rigidity dependence below and above a rigidity $P_{\mathrm{k}}$ respectively;  in this study, we set $a_2=1.95$.  The quantity $a_3=3.0$
is another dimensionless constant and determines the smoothness of the transition between
the two slopes $P^{a_1}$ and $P^{a_2}$ at $P_{\mathrm{k}}$.  The rigidity dependence is therefore essentially a double power-law.

The perpendicular diffusion coefficient $\kappa_{\bot}$ consists of perpendicular diffusion in the $\textbf{e}_r$ and $\textbf{e}_\theta$ directions respectively, i.e.
$\kappa_{\bot r}$ and $\kappa_{\bot\theta}$.  Here, $\kappa_{\bot}$ is scaled as $\kappa_{||}$.
Giacalone and Jokipii (1999) found that the ratio
$\kappa_{\bot}/\kappa_{||}$ has a value between 0.02 and 0.04.  In addition,
observations from the Ulysses spacecraft have revealed that the latitude dependence of CR protons
is significantly less than predicted by classical drift models (Potgieter and Haasbroek, 1993), which led K\'{o}ta and Jokipii (1995) to propose the concept of an anisotropic
perpendicular diffusion, where $\kappa_{\bot\theta}>\kappa_{\bot r}$ in the off-equatorial regions (e.g.  Burger, Potgieter and Heber, 2000;  Ferreira \textit{et al.}, 2000;  Potgieter, 2000).  
This anisotropy of $\kappa_{\bot}$ 
is accounted for in this work, as was done by e.g. Ngobeni and Potgieter (2008) as well as Potgieter \textit{et al.} (2014), so that
\begin{equation}
\kappa_{\bot r}=\kappa_{\bot r}^0\kappa_{||}
\end{equation}
and
\begin{equation}
\kappa_{\bot\theta}=f(\theta)\kappa_{\bot\theta}^{0}\kappa_{||},
\end{equation}
where $\kappa_{\bot r}^0=\kappa_{\bot\theta}^0=0.02$ are dimensionless constants, and
\begin{equation}
f(\theta)=A^{+}+A^{-}\tanh\left[\dfrac{1}{\Delta\theta}\left(\overset{\sim}{\theta}-\dfrac{\pi}{2}+\theta_{\mathrm{F}}\right)\right].
\end{equation}
Here $A^{\pm}=\dfrac{d\pm1}{2}$, $\Delta\theta=1/8$, with
\begin{equation}\label{eq:Apm}
\overset{\sim}{\theta}=\left\{ \begin{array}{lcl} \theta & \mathrm{for} & \theta\geq\dfrac{\pi}{2} \\ \\ \pi-\theta & \mathrm{for} & \theta < \dfrac{\pi}{2}, \end{array} \right.
\end{equation}
and
\begin{equation}
\theta_{\mathrm{F}}=\left\{\begin{array}{lcl}\dfrac{-35^{\circ}\pi}{180^{\circ}} & \mathrm{for} & \theta\geq\dfrac{\pi}{2} \\ \\ \dfrac{35^{\circ}\pi}{180^{\circ}} & \mathrm{for} & \theta < \dfrac{\pi}{2},\end{array} \right.
\end{equation}
where $d=3.0$ is a dimensionless constant that determines the enhancement factor of $\kappa_{\bot\theta}$ from its value in the equatorial plane
towards the poles, with respect to $\kappa_{||}$;  see also Potgieter (2000).

Cosmic rays will undergo a combination of gradient and curvature drifts $\langle\vec{v}_{\mathrm{d}}\rangle_{\mathrm{gc}}$ caused by
the large scale HMF, as well as current sheet drift $\left(\vec{v}_{\mathrm{d}}\right)_{\mathrm{ns}}$ because of a switch in magnetic polarity over the HCS.
These drifts are both calculated from the general equation for the pitch-angle averaged guiding center drift velocity, written as
\begin{eqnarray}\label{eq:drift_gen}
\langle\vec{v}_{\mathrm{d}}\rangle & = & \nabla\times\kappa_{\mathrm{A}}\textbf{e}_{B} \nonumber \\
							   & = & \left[\nabla\times\left(\kappa_{\mathrm{A}}\textbf{e}_B^{'}\right)\right]\left[1-2H\left(\theta-\theta^{'}\right)\right]+2\delta_{\mathrm{Dirac}}\left(\theta-\theta^{'}\right)
								      \kappa_{\mathrm{A}}\textbf{e}_B^{'}\times\nabla\left(\theta-\theta^{'}\right) \nonumber \\
							   & = & \langle\vec{v}_{\mathrm{d}}\rangle_{\mathrm{gc}}\left[1-2H\left(\theta-\theta^{'}\right)\right]+\left(\vec{v}_{\mathrm{d}}\right)_{\mathrm{ns}}\delta_{\mathrm{Dirac}}\left(\theta-\theta^{'}\right),
\end{eqnarray}
where
\begin{equation}
 \textbf{e}_B = \left[1-2H\left(\theta-\theta^{'}\right)\right]\textbf{e}_B^{'},
\end{equation}
and $\textbf{e}_B^{'}=\vec{B}/B$;  $H$ is the Heaviside function, $\delta_{\mathrm{Dirac}}$ is the Dirac delta function, $\theta^{'}$ is the HCS latitudinal extent defined below,
and $\kappa_{\mathrm{A}}$ the global drift coefficient which is given by
\begin{equation}\label{eq:ka}
 \kappa_{\mathrm{A}}=\kappa_{\mathrm{A}}^0qA\dfrac{P\beta}{3B}\dfrac{\left(P/P_0\right)^2}{1+\left(P/P_0\right)^2}.
\end{equation}
In this equation $q$ is the particle charge sign so that in this study, which only considers results pertaining to CR proton modulation, $q=+1$.  
The quantity $P_0\in\lbrace1/\sqrt{10},1/\sqrt{40}\rbrace$ GV is added on dimensional grounds.
The value $A=\pm1$, i.e. either $A>0$ or $A<0$;  this represents the magnetic polarity cycle with $A>0$ ($A<0$) defined as 
the polarity cycle in which the HMF points outward (inward) in the Northern hemisphere of the Sun;  during an $A>0$ ($A<0$) polarity cycle positively charged
particles will drift inward (outward) along the polar regions and outward (inward) along the HCS in the equatorial regions, while the reverse is true for negatively charged particles.
The quantity $\kappa_{\mathrm{A}}^0$ is a constant which can be used as a scaling factor, having values between 0.0 (zero drift) and 1.0 (full drift);  
in this work, $\kappa_{\mathrm{A}}^0=1.0$, which is known as the so-called weak scattering maximal value for the drift coefficient.  Remembering the change in sign that occurs
when crossing the HCS, the gradient and curvature drifts can thus be calculated by means of
\begin{eqnarray}
 \langle\vec{v}_{\mathrm{d}}\rangle_{\mathrm{gc}}  & = &\nabla\times\left(\kappa_{\mathrm{A}}\dfrac{\vec{B}}{B}\right) \nonumber \\
										       & = & \kappa_{\mathrm{A}}^0qA\dfrac{ P \beta}{3}\dfrac{\left(P/P_0\right)^2}{1+\left(P/P_0\right)^2}\nabla\times\left(\dfrac{\vec{B}}{B^2}\right)\\ \nonumber
										       & =& \kappa_{\mathrm{A}}^{'}\nabla\times\left(\dfrac{\vec{B}}{B^2}\right),
\end{eqnarray} 
where $\kappa_{\mathrm{A}}^{'}$ was defined as
\begin{equation}
 \kappa_{\mathrm{A}}^{'}=\kappa_{\mathrm{A}}^0qA\dfrac{P\beta}{3}\dfrac{\left(P/P_0\right)^2}{1+\left(P/P_0\right)^2}.
\end{equation}

\begin{figure}
\begin{center}
\includegraphics[height=5.0 in,width=3.8 in, angle = 90]{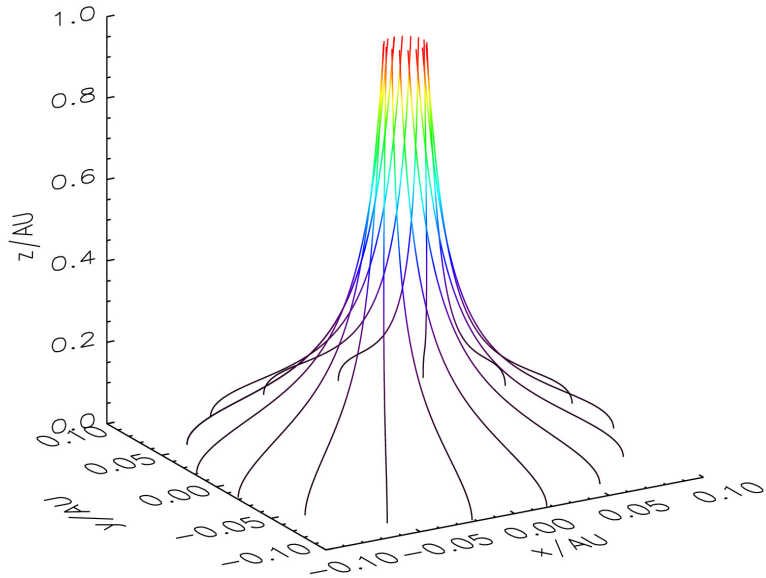}
\caption{\small \sl Gradient and curvature drift streamlines in the case
of the Parker HMF, shown in Cartesian coordinates.  From the narrow, cone-like configuration of these lines it is evident that drift through the polar regions of the heliosphere proceeds effectively.
The Sun is again located at $\left(X,Y,Z\right)=\left(0,0,0\right)$.  The colour scale corresponds to the magnitude of the gradient and curvature drift velocity.}\label{fig:PHMFdrift}
\end{center}
\end{figure}

For the Parker HMF, the averaged gradient and curvature drift velocities in the radial, latitudinal, and azimuthal directions respectively are given by
\begin{eqnarray}\label{eq:PHMFdrift}
v_{\mathrm{dr}} & = & -v_{\mathrm{d}0}\cot\theta \nonumber \\ 
v_{\mathrm{d\theta}} & = & v_{\mathrm{d}0}\left(2+\Psi^2\right)\nonumber \\ 
v_{\mathrm{d\phi}} & = & v_{\mathrm{d}0}\Psi\cot\theta,
\end{eqnarray}
where
 \begin{equation}
v_{\mathrm{d}0}=2\dfrac{\kappa_{\mathrm{A}}^{'}\Psi}{rB\sqrt{1+\Psi^2}^3}
 \end{equation}
Figure \ref{fig:PHMFdrift} shows the drift velocity streamlines for this case, i.e. solving for 
\begin{equation}\label{eq:streamlines}
 \dfrac{d\vec{r}(l)}{dl}=\langle\vec{v}_{\mathrm{d}}\rangle_{\mathrm{gc}},
\end{equation}
where $\vec{r}=\left(r,\theta,\phi\right)$ is the position vector in spherical coordinates and $l$ is the arc length along the field lines of $\langle\vec{v}_{\mathrm{d}}\rangle_{\mathrm{gc}}$.  
These lines are calculated in the polar regions and it is clear from their narrow, cone-like configuration that drift in these regions occurs very effectively,
towards the Sun.  Note that Figure \ref{fig:PHMFdrift} again makes use of Cartesian coordinates so that the Sun is located at $\left(X,Y,Z\right)=\left(0,0,0\right)$;  the polar
regions are centered around the $Z$-axis.

When accounting for current sheet drift, which is directed parallel to the HCS and perpendicular to the HMF (e.g. Burger, Moraal and Webb, 1985;  Burger and Potgieter, 1989),
the model does not make use of the $\delta_{\mathrm{Delta}}$ function in Eq. (\ref{eq:drift_gen});  rather, it is
assumed that a particle will experience current sheet drift upon the condition that its distance $d_{\mathrm{ns}}$ to the HCS is less than two gyro radii, i.e.
\begin{equation}\label{eq:HCScondition}
 d_{\mathrm{ns}}\leq2r_{\mathrm{L}}=2\left[\dfrac{mv}{q^{'}B}\right],
\end{equation}
with $q^{'}$ the particle charge (not to be confused with the particle charge sign $q$ as in Eq. (\ref{eq:ka})), and $m$ is the relativistic mass of the particle.
Assuming that the latitudinal extent of the HCS is given by the expression from  K\'{o}ta and Jokipii (1983)
\begin{equation}
 \theta^{'}=\dfrac{\pi}{2}-\arctan\left[\tan\alpha\sin\left(\phi+\dfrac{\Omega(r-r_\odot)}{V_{\mathrm{sw}}}\right)\right],
\end{equation}
Eq. (\ref{eq:PHMFdrift}) is replaced by
\begin{eqnarray}\label{eq:HCSdrift}
v_{\mathrm{d}r} & = & v_{\mathrm{ns},0}\sin\psi\cos\left(\pm\beta_{\mathrm{rot}}\right) \nonumber \\
v_{\mathrm{d}\theta} & = & v_{\mathrm{ns},0}\sin\left(\pm\beta_{\mathrm{rot}}\right) \nonumber \\
v_{\mathrm{d}\phi} & = & v_{\mathrm{ns},0}\cos\psi\cos\left(\pm\beta_{\mathrm{rot}}\right),
\end{eqnarray}
when the condition of Eq. (\ref{eq:HCScondition}) is met.   Here $v_{\mathrm{ns},0}$ is given by the approximation of Burger, Moraal and Webb (1985)
\begin{equation}
 v_{\mathrm{ns},0}=vqA\left[0.457-0.412\dfrac{d_{\mathrm{ns}}}{r_\mathrm{L}}+0.0915\left(\dfrac{d_{\mathrm{ns}}}{r_{\mathrm{L}}}\right)^2\right],
\end{equation}
and $\beta_{\mathrm{rot}}$ is the angle between the radial direction and a vector parallel to $\left(\vec{v}_{\mathrm{d}}\right)_{\mathrm{ns}}$, defined by
\begin{equation}
 \tan\beta_{\mathrm{rot}}=r\dfrac{\partial\theta^{'}}{\partial r},
\end{equation}
as in Strauss \textit{et al.} (2012) and discussed by Burger (2012).  The correct sign of $\beta$ in Eq. (\ref{eq:HCSdrift}) is determined by
\begin{equation}
\begin{array}{lcl} \dfrac{\partial\theta^{'}}{\partial r}<0&\Rightarrow&\beta_{\mathrm{rot}} <0 \\ \\
\dfrac{\partial\theta^{'}}{\partial r}\geq0&\Rightarrow&\beta_{\mathrm{rot}} >0.
\end{array}
\end{equation}

The proton LIS assumed in this study has its origin in that of Langner and Potgieter (2004).  Potgieter \textit{et al.} (2014) improved on this LIS by taking into 
account PAMELA measurements at energies between 30 and 50 GeV.  However, this LIS did not provide for the heliopause (HP) crossing of Voyager 1 and was therefore 
modified to take into account these new Voyager 1 observations in the lower energy range;  see Potgieter (2014) in this regard.  The LIS  resulting from these modifications,
expressed in units of $\mathrm{particles.m^{-2}.s^{-1}.sr^{-1}.MeV^{-1}}$, is employed in this study and is given by the expressions
\begin{center}
 $j=0.6978\exp\left\{ 4.64 - 0.023(\ln E)^2 - 2.91E^{-0.5}\right\} $
\end{center}
for energies $E<1.4$ GeV, and
\begin{equation}
 j=0.6847\exp\left\{3.22 - 2.78(\ln E) - 1.5E^{-1}\right\}
\end{equation}
for $E>1.4$ GeV.  The HP position is assumed at 120 AU, where this LIS is specified as an initial condition for cosmic ray modulation.

\section{The Numerical Modulation Model}

Since analytical solutions to the TPE in Eq. (\ref{eq:TPE}) are only possible to a very limited extent, modulation models solve this equation numerically to various degrees of complexity.  
The numerical model of this work makes use of an SDE approach and is based on the well-benchmarked model of Strauss \textit{\textit{et al.}} (2011a, 2011b, 2012).  
It is only briefly explained here;  an extensive treatment can be found in Kopp \textit{et al.} (2012).  The model is solved in a steady state and assuming no sources so that $Q=0$.
It is solved in spherical coordinates $(r,\theta,\phi)$ and particle kinetic energy $E$.

Each Fokker-Planck type equation has a corresponding set of independent SDEs, which is calculated in either a time forward or time backward fashion.
The TPE, written in spherical coordinates, is associated with such a time backward set of SDEs, given by
\begin{equation}\label{eq:SDEs}
\begin{array}{lcl} dr & = & \left[\dfrac{1}{r^2}\dfrac{\partial}{\partial r}\left(r^2\kappa_{rr}\right)+\dfrac{1}{r\sin\theta}\dfrac{\partial\kappa_{r\phi}}{\partial\phi}-V_{\mathrm{sw}}-v_{dr}\right]ds + \sqrt{2\kappa_{rr}-\dfrac{2\kappa^2_{r\phi}}{\kappa_{\phi\phi}}}dW_r+\dfrac{\sqrt{2}\kappa_{r\phi}}{\sqrt{\kappa_{\phi\phi}}}dW_{\phi} \\ \\ 
d\theta & = & \left[\dfrac{1}{r^2\sin\theta}\dfrac{\partial}{\partial \theta}\left(\sin\theta\kappa_{\theta\theta}\right)-\dfrac{v_{d\theta}}{r}\right]ds + \dfrac{\sqrt{2\kappa_{\theta\theta}}}{r}dW_{\theta} \\ \\
d\phi & = & \left[\dfrac{1}{r^2\sin^2\theta}\dfrac{\partial\kappa_{\phi\phi}}{\partial\phi}+\dfrac{1}{r^2\sin\theta}\dfrac{\partial}{\partial r}\left(r\kappa_{r\phi}\right)-\dfrac{v_{d\phi}}{r\sin\theta}\right]ds + \dfrac{\sqrt{2\kappa_{\phi\phi}}}{r\sin\theta}dW_{\phi} \\ \\
dE & = & \left[\dfrac{1}{3r^2}\dfrac{\partial}{\partial r}\left(r^2V_{\mathrm{sw}}\right)\Gamma E\right] ds,\end{array}
\end{equation}
for each of the spherical coordinates $r$, $\theta$, and $\phi$, as well as for the particle energy $E$;  $d\vec{W}=\left[dW_r,dW_{\theta},dW_{\phi}\right]$ is the multi-dimensional Wiener 
process with each of its elements containing a Gaussian distributed random number.  The quantity $\Gamma$ is defined by
\begin{equation}
\Gamma=\dfrac{E+2E_0}{E+E_0}
\end{equation}
where $E_0$ is the particle rest energy; $ds$ is the infinitesimal backward time increment.

The diffusion tensor has here been converted to spherical coordinates and also includes the drift coefficient $\kappa_{\mathrm{A}}$.  Its elements are related to 
the coefficients in HMF aligned coordinates through
\begin{equation}\label{eq:Kspher}
\begin{array}{lcl}
\left[{\begin{array}{ccc}\kappa_{rr}&\kappa_{r\theta}&\kappa_{r\phi}\\ \kappa_{\theta r} & \kappa_{\theta\theta} & \kappa_{\theta\phi} \\
	      \kappa_{\phi r}&\kappa_{\phi\theta}&\kappa_{\phi\phi}\end{array}}\right] 
&=&\left[{\begin{array}{ccc} \kappa_{||}\cos^2\psi+\kappa_{\bot r}\sin^2\psi & -\kappa_{\mathrm{A}}\sin\psi & (\kappa_{\bot r}-\kappa_{||})\cos\psi\sin\psi \\ 
	      \kappa_{\mathrm{A}}\sin\psi & \kappa_{\bot\theta} & \kappa_{\mathrm{A}}\cos\psi \\ 
	      (\kappa_{\bot r}-\kappa_{||})\cos\psi\sin\psi & -\kappa_{\mathrm{A}}\cos\psi & \kappa_{||}\sin^2\psi+\kappa_{\bot r}\cos^2\psi \end{array}}\right].\\ \\
\end{array}
\end{equation}
This tensor transformation is only valid for Parker-like HMF profiles, i.e. for an HMF that does not contain a $\theta$ component;  more general transformations are
discussed in e.g. Effenberger \textit{et al.} (2012).

Solving the TPE in a time backward fashion implies that a number of pseudo-particles, $N$, are traced from Earth outward to the boundary of the heliosphere,
stepping in $(r,\theta,\phi)$ and calculating energy losses according to Eq. (\ref{eq:SDEs}).  At the outer modulation boundary, $r_{\mathrm{out}}=120$ AU,
particle relative intensities are convolved with the LIS and an energy spectrum at Earth is obtained.  For the purposes of this study, 
no modulation is assumed beyond the HP, so that the heliopause spectrum and the very LIS is synonymous (see Potgieter \textit{et al.}, 2013). 
A reflective inner modulation boundary is used at $r_{\mathrm{in}}=0.1$ AU, so that any pseudo-particle for which $r<r_{\mathrm{in}}$ is reflected
back into the modulation volume, i.e. $r_{\mathrm{in}}<r<r_{\mathrm{out}}$.

\section{Modified Magnetic Field Profiles}
\label{sec:HMFmod}

\subsection{The modification of Jokipii and K\'{o}ta}
Jokipii and K\'{o}ta (1989) showed that the effect of small perpendicular magnetic field components
near the solar surface leads to a larger magnetic field at greater radial distances in the polar regions.  These authors
argued that the average direction of these small components would cancel out so that only the
magnitude of the magnetic field is modified.  They put forward the following modified expression for the Parker HMF magnitude
\begin{equation}\label{eq:JKM}
 B=B_0\left[\dfrac{r_0}{r}\right]^2\sqrt{1 + \Psi^2 + \left( \dfrac{r\delta }{r_{\odot}} \right)^2},
\end{equation}
introducing the quantity $\delta$ to signify the magnitude of the extra, superimposed, transverse magnetic field.  The Jokipii-K\'{o}ta modification is henceforth
referred to as the JKM, while the term containing $\delta$ will be referred to as the
JKM modification term.  In this study we use $\delta=0.002$ (see e.g. Potgieter, 2000).  Geometrically speaking, a cone centered around the polar regions is induced in which 
the HMF is altered so that it decreases as $r^{-1}$ instead of $r^{-2}$.  The effect of this modification, therefore, is to bring about the changes which 
are required to reduce the large drift velocities in the polar regions, without noticeably altering the field in the equatorial plane.

However, upon this
modification, the requirement that the magnetic field remains divergence free, i.e. $\nabla\cdot\vec{B} = 0 $, is no longer complied with and therefore the JKM is technically incorrect.  
Through inspection, the HMF will remain divergence free if $\delta=\delta(\theta)\propto\left(\sin\theta\right)^{-1}$, so that
\begin{equation}\label{eq:JKMmod}
 \delta (\theta) = \dfrac{\delta_{\mathrm{m}}}{\sin\theta},
\end{equation}
with $\delta_{\mathrm{m}}=8.7\times 10^{-5}$ (Langner, 2004).  This value for $\delta_{\mathrm{m}}$ implies $\delta (\theta) = 0.002$ 
near the poles and $\delta (\theta) \approx 0 $ in the ecliptic plane.  This will again bring about the intended changes to the HMF.

\subsection{The modification of Smith and Bieber}
This modification was introduced by Smith and Bieber (1991), who analysed data collected by various satellites and found magnetic field spirals
at Earth more tightly wound than that predicted by the Parker theory.  Consequently, they suggested
a modification to the Parker spiral by reasoning that the difference in rotational speed of the equatorial 
and polar regions of the Sun would cause small azimuthal magnetic field components to develop.  This would 
lead to larger spiral angles at larger radial distances, and would provide an explanation for the larger
spiral angles at Earth.  At the same time, this modification would have a significant effect on the HMF
structure at the high latitude regions.  
The expression for the spiral angle, i.e. Eq. (\ref{eq:PHMFspiral}), was modified to yield
\begin{equation}\label{eq:SBM}
 \Psi^{'}=\dfrac{\Omega(r-b)\sin\theta}{V_{\mathrm{sw}}(r,\theta)}-\dfrac{r}{b}\dfrac{V_{\mathrm{sw}}(b,\theta)}{V_{\mathrm{sw}}(r,\theta)}\left(\dfrac{B_{\mathrm{T}}(b)}{B_{\mathrm{R}}(b)}\right),
\end{equation}
where $B_{\mathrm{T}}(b)/B_{\mathrm{R}}(b)$ is the ratio of the azimuthal to the radial magnetic field components at a
position $b$ near the solar surface, here taken as $b=20r_{\odot}$.  Smith and Bieber (1991) showed that the value of 
$B_{\mathrm{T}}(b)/B_{\mathrm{R}}(b)$ is approximately -0.02 and although the value of $b$ assumed here is larger than
the original $b=5r_{\odot}$ assumed by these authors, the value of -0.02 is retained for the purposes of this paper, except where
explicitly indicated otherwise.  This modification, based on sound physical arguments, has been mostly ignored in the numerical 
modeling of the solar modulation of cosmic rays, which is rather surprising. It will be shown next that this modification is indeed 
appropriate and as such an improvement.

\begin{figure}
\begin{center}
\includegraphics[height=5.0 in,width=3.8 in, angle = 90]{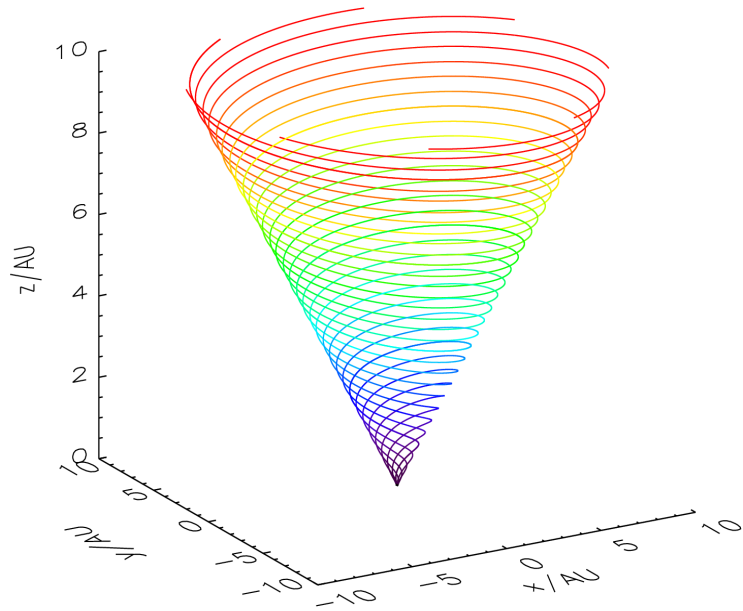}
\caption{\small \sl Magnetic field lines resulting from employing the SBM with $B_{\mathrm{T}}(b)/B_{\mathrm{R}}(b)=-0.5$.  Compared to Figure \ref{fig:PHMF}, these lines are evidently
closer together than in the case of the unmodified Parker HMF, in accordance with the larger HMF magnitude of the modified field.  The field lines again start from
successive azimuthal angles, latitude $\theta=45^\circ$, and is shown over the first 10 AU from the Sun;  the colour scale again corresponds to $r^2|B|$.}\label{fig:SBM}
\end{center}
\end{figure}

Figure \ref{fig:SBM} is an illustration of HMF lines that results from applying the SBM;  these lines once again originate at consecutive azimuthal angles
and at a latitude of $\theta=45^{\circ}$.
According to Eq. (\ref{eq:SBM}), and clearly evident from comparing Figures \ref{fig:SBM} and \ref{fig:PHMF}, the SBM 
alters the direction of the Parker HMF by winding up the magnetic field spirals
more tightly; as a consequence, in contrast to the straight line case of Eqs. (\ref{eq:PHMFspiral}) and (\ref{eq:PHMF}),
the HMF still has an azimuthal component in the polar regions, even at $\theta=0^{\circ}$.  It is also clear from this comparison
that the field lines in the case of the SBM are closer together than in the case of the Parker HMF, in accordance with the larger HMF magnitudes of the modified field.
Note that, in Figure \ref{fig:SBM}, $B_{\mathrm{T}}(b)/B_{\mathrm{R}}(b)=-0.5$ was chosen to be very small, corresponding to a very large modification, in order to illustrate these
effects more clearly.

It should be noted that the ultimate, very complex, HMF modification was introduced by Fisk (1996). However,  it has been seen as quite controversial and 
without convincing observational support so that it is not pursued any further in this work;  see Sternal \textit{et al.} (2011) for an appreciation of this modification. 
The SDE-approach does however allow for the implementation of a HMF of this complexity whereas traditional ADI based numerical schemes 
are quite unsuitable.

\subsection{Comparison of the features of HMF modifications}
\begin{figure}
 \begin{center}
\includegraphics[height=9.0 in,width=6.5 in, angle = 0]{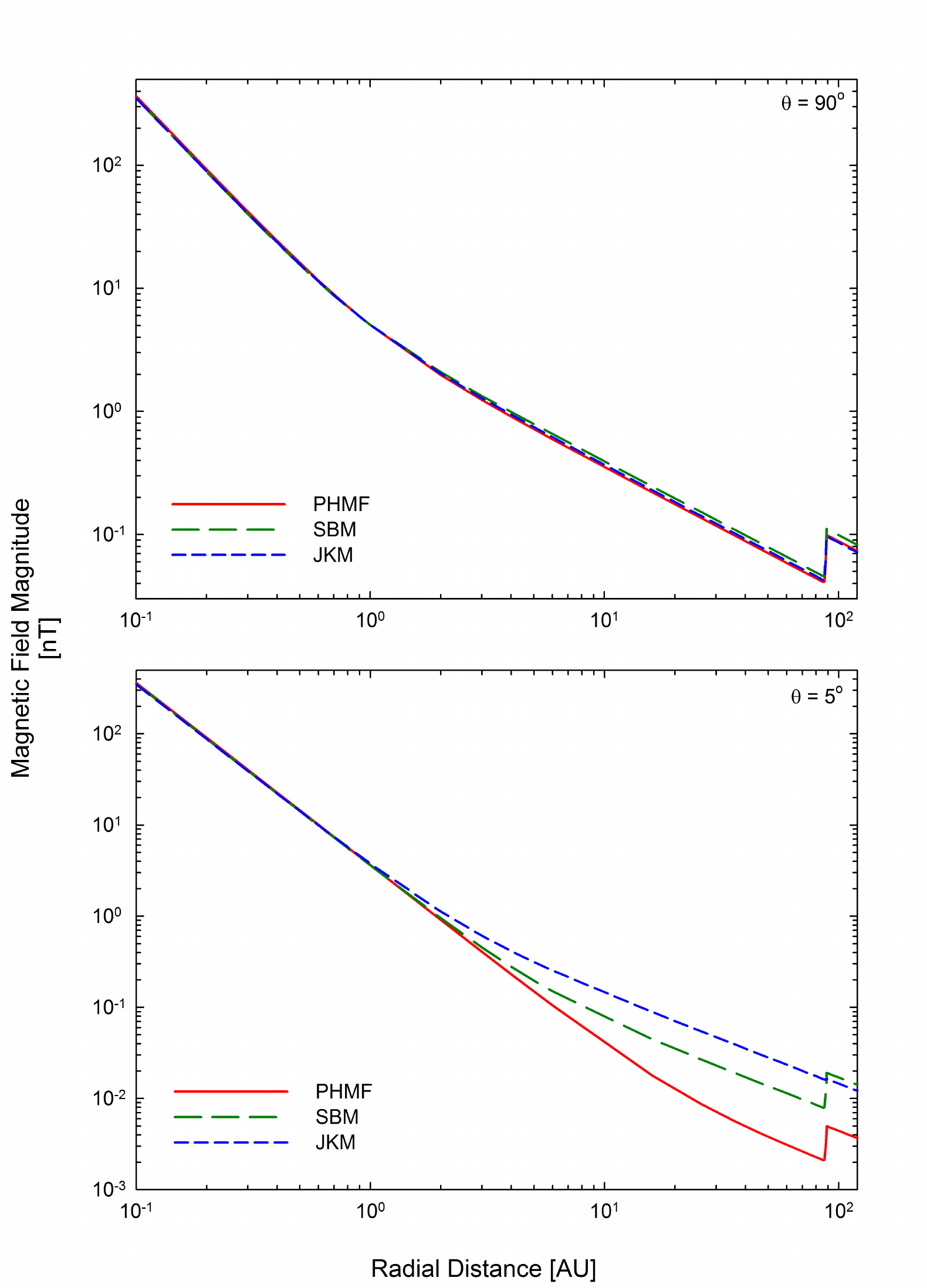}
\caption{\small \sl Magnetic field magnitude radial profiles in the equatorial regions (top panel) and in the polar regions (bottom panel).  The Parker HMF is indicated by the solid red line while 
the SBM and JKM are represented by the dashed green and blue lines respectively.}\label{fig:Brad}
 \end{center}
\end{figure}

The top panel of Figure \ref{fig:Brad} compares the radial profiles of the Parker HMF and HMFs which have been modified according to respectively the SBM and JKM;  this is shown
in the equatorial plane with $\theta=90^\circ$, assuming a magnetic field magnitude of 
5.05 nT at Earth.  All three of these HMF profiles start out with an $r^{-1.95}$ dependence at the smallest radial distances and end with an $r^{-1.00}$ 
dependence just in front of the TS at $r=r_{\mathrm{TS}}$, where $V_{\mathrm{sw}}$ drops by a factor of 2.5 and brings about
a step-like increase;  beyond the TS, the $r^{-1.00}$ dependent decrease is continued.  In the case of the Parker HMF, the $r^{-1.95}$ dependence 
continues up to a radial distance of about 0.5 AU, where it begins to transition into a weaker dependence to eventually reach $r^{-1.00}$ at $\sim$ 3.0 AU.  
Although not clearly discernible on the scale of this graph,  the SBM and JKM profiles were determined to enter the $r^{-1.00}$ dependence by a distance of up to 1 
AU earlier than the Parker HMF.

The bottom panel of Figure \ref{fig:Brad} compares the radial profiles of the Parker HMF, SBM, and JKM in the polar regions and necessarily illustrates the effects of the modifications more clearly.  
The Parker HMF and both the SBM and the JKM 
start out with an $r^{-1.96}$ dependence at the smallest radial distances.  At a radial distance between 6 AU and 7 AU the Parker HMF starts transitioning into a weaker 
radial dependence and keeps up this transition to reach its weakest dependence of $r^{-1.10}$ at a radial distance of between 40 AU and 50 AU; 
this is then maintained up to the TS.  The SBM begins its transition into the weaker radial dependence at a smaller radial distance between 1 AU and 2 AU and 
reaches $r^{-1.00}$ at about 20 AU, continuing up to the TS.  The JKM begins the transition into the weaker radial dependence even earlier than the SBM at $\sim$ 1 AU 
and reaches $r^{-1.00}$ between 8 and 9 AU, once again continuing up to the TS.  In all three cases, the dependence in front of the TS is regained beyond the TS.  

Notice that, in the polar regions, the JKM profile does not undergo the upward step at the TS to the extent shown by the Parker HMF and SBM;  in fact, the JKM profile barely shows 
any visible increase.  This suppression of the TS in the polar regions is not an intended effect and is a consequence of the fact that the TS is simulated entirely by the drop in $V_{\mathrm{sw}}$
at $r=r_{\mathrm{TS}}$.  In the instance of the JKM, the modification term in Eq. (\ref{eq:JKM}) is much larger than the term containing $V_{\mathrm{sw}}$ and hence the effect of the 
drop in $V_{\mathrm{sw}}$ at $r=r_{\mathrm{TS}}$ is markedly less pronounced.

It is clear from Figure \ref{fig:Brad} that both the SBM and JKM have the desired effect of drastically reducing the radial dependence of the HMF in the polar regions:  the transition 
into the weaker radial dependence is initiated at much smaller $r$ than in the case of the Parker HMF, ensuring significantly larger magnetic field magnitudes $B$.  
It is also noted that because the SBM starts its transition into the weaker radial dependence at larger radial distances than does the JKM, its effect on $B$ is less pronounced than that of the 
JKM.  Although not illustrated in Figure \ref{fig:Brad}, it is of importance to note that this less effectual status of the SBM when compared to the JKM is only \textit{locally} 
true, i.e. it is only valid at or \textit{very close} to the poles as in this case with $\theta=5^{\circ}$.

\begin{figure}
\begin{center}
\includegraphics[height=9.0 in,width=4.8 in, angle = 0]{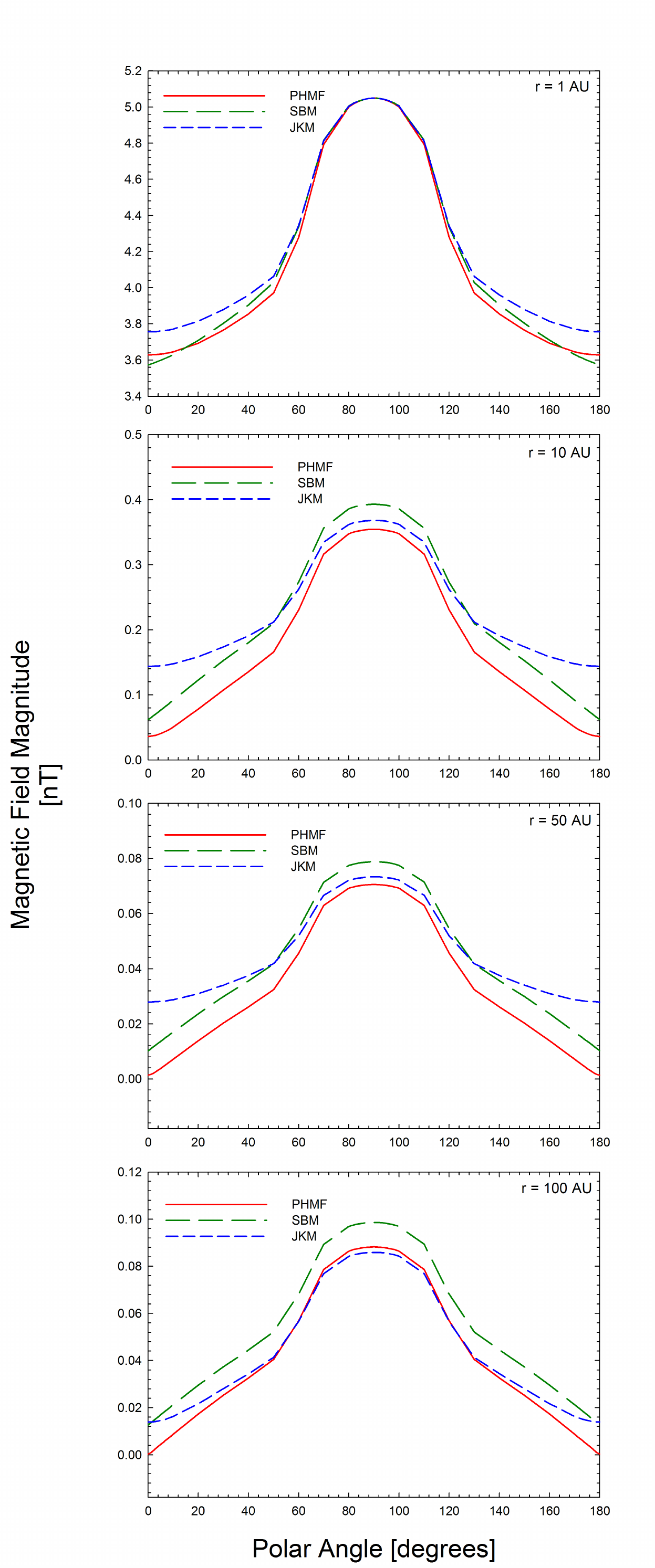}
\caption{\small \sl Magnetic field latitudinal profiles at four selected radial distances $r=1$ AU (first panel), 10 AU (second panel), 50 AU (third panel), and 100 AU (fourth panel).  
The Parker HMF is indicated by the solid red line while the SBM and JKM are represented by the dashed green and blue lines respectively.}\label{fig:Blat}
\end{center}
\end{figure}

The four panels of Figure \ref{fig:Blat} show the unfolding of the Parker HMF, SBM, and JKM latitudinal profiles over successive radial distances of 1 AU, 10 AU, 50 AU, and 100 AU.  From 
the first panel it is clear that, at 1 AU, the effects of both the SBM and JKM are negligible in the equatorial regions, but become more pronounced towards the poles.  Taking
the equator at $\theta=90^{\circ}$ as reference and moving towards the poles, the deviation of the SBM from the Parker HMF is seen to start at about $\theta=90^\circ\pm22^\circ$,
continuing towards the respective poles.  The SBM is larger than the Parker HMF up to about $\theta=90^\circ\pm76^\circ$, where it drops below and eventually
reaches a maximum deviation from the Parker HMF of $\sim$ 0.08 nT at the poles.  The difference between the Parker HMF and JKM is seen to consistently increase towards the polar 
regions $\theta\in \lbrace 0^\circ,180^\circ\rbrace$, where the JKM eventually reaches levels of $\sim$ 0.12 nT higher than the Parker HMF. 

The picture at a radial distance of 10 AU has unfolded to be significantly different from the profiles at 1 AU.  The SBM is seen to have already modified the Parker HMF in the 
equatorial regions, amounting to a value of about 0.04 nT higher than the Parker HMF at $\theta = 90^\circ$.  Towards the poles the difference between the SBM and Parker HMF increases 
to about 0.06 nT at intermediate latitudes, declining again to about 0.03 nT at $\theta\in\lbrace 0^\circ, 180^\circ \rbrace$.  The SBM, at the equator, is initially higher than 
the JKM, but drops below it at about $\theta=90^\circ\pm42^\circ$, and continues to do so up to the polar regions.  Smaller deviations from the Parker HMF in the 
equatorial regions are also seen to be induced by the JKM, amounting to no more than 0.02 nT at $\theta = 90^\circ$.  As was the case at 1 AU, the effect of the JKM 
increases towards the poles where it reaches levels of about 0.10 nT higher than the Parker HMF.   The picture at 50 AU looks qualitatively the same as that at 10 AU, only the 
magnitude of the HMF being smaller.  Comparing with Figure \ref{fig:Brad}, it is clear that this qualitative picture is continued up to the TS.

The fourth panel of Figure \ref{fig:Blat} shows the latitudinal profiles at 100 AU, a distance comfortably beyond the TS.  The SBM is still higher than the Parker HMF at all latitudes, now
being higher than the Parker HMF by about 0.01 nT in the equatorial regions and by about 0.02 nT in the polar regions.  The JKM is, curiously enough, at more or less the same
level as the SBM in the polar regions;  it lies just below the Parker HMF at $\theta = 90^\circ$, but climbs above it at about $\theta=90^\circ\pm45^\circ$, eventually
reaching the level of the SBM at the poles.  This unfolding of the HMF latitudinal profiles over successive radial distances effectively illustrates the previous statement about
the more effective status of the JKM being only locally true:  the SBM is effected over a larger part
of the modulation volume so that its overall effect can indeed be greater than that of the JKM.

\begin{figure}
\begin{center}
\includegraphics[height=3.6 in,width=5.0 in, angle = 0]{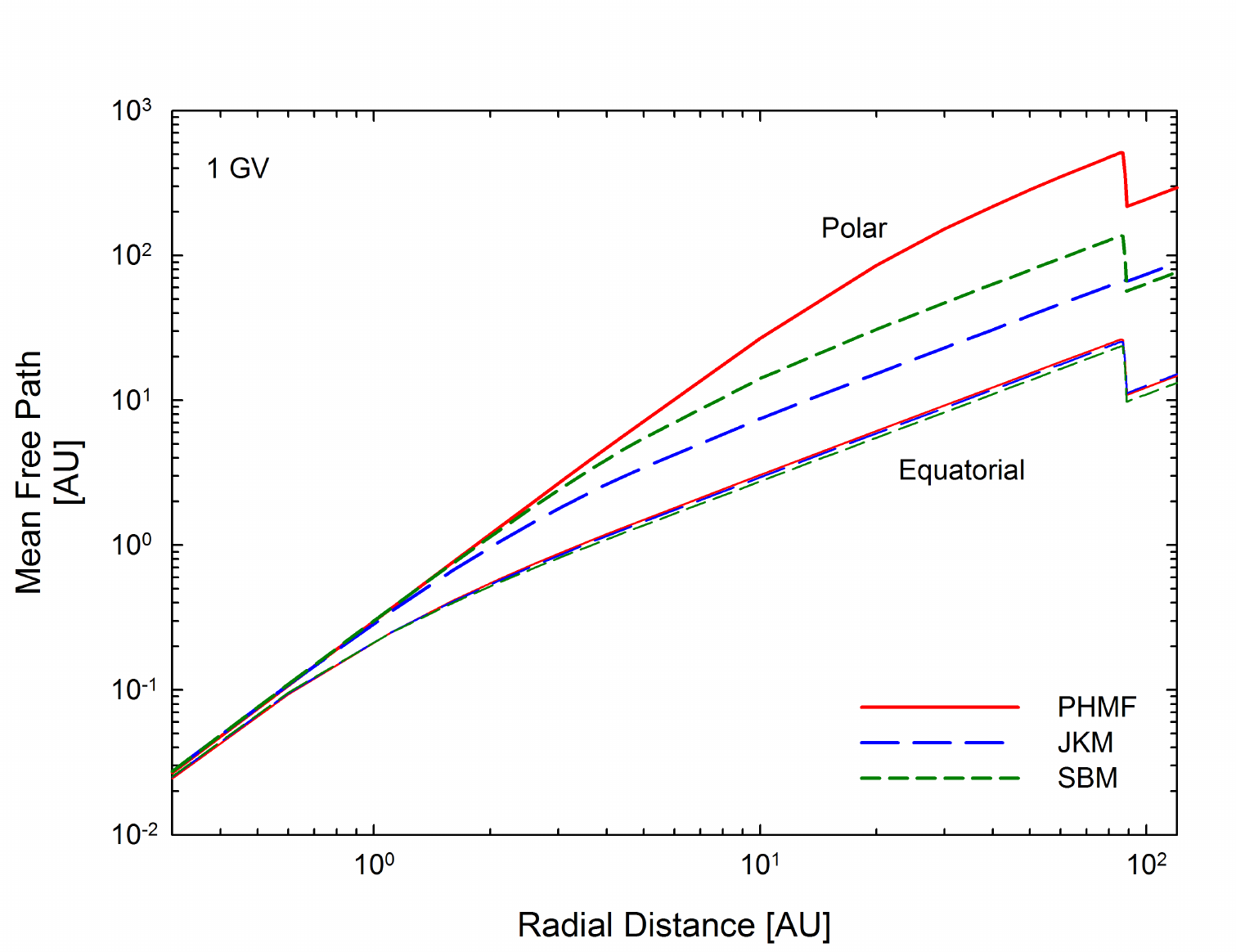}
\caption{\small \sl Radial profiles for the parallel mean free paths $\lambda_{||}$ of a 1 GV proton for each of the Parker HMF (red lines), the SBM (green lines), and the JKM (blue lines).  
The upper three lines depict the $\lambda_{||}$s in the polar regions, while the lower three lines do so in the equatorial regions.}\label{fig:lambda}
\end{center}
\end{figure}

From Eqs.(\ref{eq:kpar}) to (\ref{eq:Apm}), defining the diffusion coefficients straight-forwardly as used in this study, it is clear that the choice of HMF profile directly
influences the spatial dependence of these coefficients.  
Figure \ref{fig:lambda} shows the radial profiles of the parallel mean free paths $\lambda_{||}$ of 1 GV protons, which is related to the parallel diffusion coefficient via
\begin{equation}
\lambda_{||}=\dfrac{3}{v}\kappa_{||}.
\end{equation}
This is shown for each of the Parker HMF
(solid red lines), the SBM (dashed green lines), and the JKM (dashed blue lines).  The upper three lines represent the values of $\lambda_{||}$ in the polar ($\theta=5^\circ$) regions, while the 
lower three, almost coinciding, lines do so in the equatorial plane at $\theta=90^\circ$.  As expected, the effects of the modifications are minimal in the equatorial regions, while being far more significant in the polar regions.  
In the equatorial regions, the $\lambda_{||}$s for each of the Parker HMF, SBM and JKM are identical at small $r$ and then start diverging from one another at about 1 AU.  This divergence
is not significant however, and both the $\lambda_{||}$s for the SBM and JKM end up being only marginally below the Parker HMF, the SBM being the lowest.
In the polar regions, the $\lambda_{||}$s for each of the Parker HMF, SBM and JKM are again seen to start out together at small $r$
and then to diverge at greater $r$.  As should be the case, the $\lambda_{||}$s for the SBM and JKM are observed to start diverging from the Parker HMF at more or less the same
radial distances at which the HMF profiles in Figure \ref{fig:Brad} started diverging from one another. In the polar regions the $\lambda_{||}$s for both the SBM and JKM then keep significantly lower
than that for the Parker HMF, the JKM being the lowest up to $r=r_{\mathrm{TS}}$ where the SBM drops below it due to the failure of the JKM to decrease over the TS.

The drift coefficient also has a $1/B$ dependence according to Eq. (\ref{eq:ka}), and a modification to the HMF will therefore
affect the particle drift velocity, so that in each of the cases for the JKM and SBM  a set of expressions analogous to Eq. (\ref{eq:PHMFdrift}) can be obtained.
For the modified JKM, i.e. using Eq. (\ref{eq:JKMmod}), the drift velocities of Eq. (\ref{eq:PHMFdrift}) are replaced by
\begin{eqnarray}\label{eq:JKMdrift}
v_{\mathrm{d}r} & = & -v_{\mathrm{d}0}\left(1+\left(\dfrac{\delta r}{r_\odot}\right)^2\right)\Psi\cot\theta \nonumber \\ 
v_{\mathrm{d}\theta}  & = & v_{\mathrm{d}0}\left(2+\left(\dfrac{\delta r}{r_\odot}\right)^2+\Psi^2\right)\Psi\nonumber \\ 
v_{\mathrm{d}\phi} & = & v_{\mathrm{d}0}\left(\Psi^2\cot\theta+\dfrac{\delta r}{r_\odot}\left(2+\left(\dfrac{\delta r}{r_\odot}\right)^2+\Psi^2\right)\right),
\end{eqnarray}
with
\begin{equation}\label{eq:JKMkappa}
v_{\mathrm{d}0}=2\dfrac{\kappa_{\mathrm{A}}^{'}}{rB\sqrt{1+\left(\dfrac{\delta r}{r_\odot}\right)^2+\Psi^2}^3}.
\end{equation}

To obtain the analogous set of equations in the case of an HMF modified according to the SBM, we write
\begin{equation}
 \Psi^{'}\approx\Psi-\delta^{'}r,
\end{equation}
where the SBM modification term $\delta^{'}$ is defined as
\begin{equation}
 \delta^{'}=-\dfrac{0.02}{b}\dfrac{V_{\mathrm{sw}}\left(b,\theta\right)}{V_{\mathrm{sw}}\left(r,\theta\right)},
\end{equation}
since we have set $B_{\mathrm{T}}(b)/B_{\mathrm{R}}(b)=-0.02$ for our choice of $b$;  finally, substituting
\begin{equation}
 \gamma = r\delta^{'} - \Psi,
\end{equation}
Eq. (\ref{eq:PHMFdrift}) is replaced by
\begin{eqnarray}
v_{\mathrm{d}r} & = & v_{\mathrm{d}0}\left(\gamma^3+\gamma^2\Psi + \gamma-\Psi\right)\cot\theta \nonumber \\ 
v_{\mathrm{d}\theta}  & = & -2v_{\mathrm{d}0}\left(2+\gamma^2\right)\gamma\nonumber \\ 
v_{\mathrm{d}\phi} & = & -2v_{\mathrm{d}0}\gamma\Psi\cot\theta,
\end{eqnarray}
with
\begin{equation}
v_{\mathrm{d}0}=\dfrac{\kappa_{\mathrm{A}}^{'}}{rB\sqrt{1+\gamma^2}^3}.
\end{equation}

\begin{figure}
\begin{center}
\includegraphics[height=5.0 in,width=3.8 in, angle = 90]{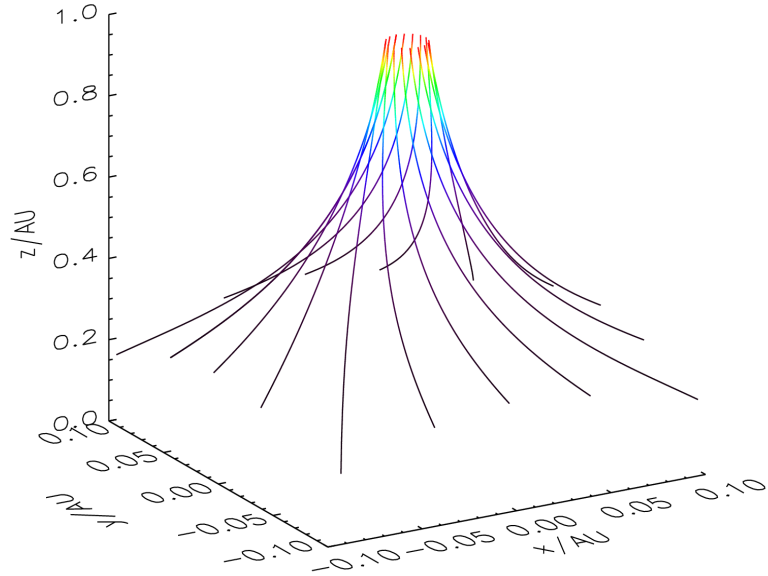}
\caption{\small \sl Gradient and curvature drift streamlines, upon employing the SBM with $B_{\mathrm{T}}(b)/B_{\mathrm{R}}(b)=-0.1$.  Compared to Figure \ref{fig:PHMFdrift},
the cone-like configuration of these lines are far less narrow, i.e. it is significantly more fanned out.  This indicates less effective inward drift
than in the case of the Parker HMF.  Once again, the colour scale corresponds to the magnitude of the gradient and curvature drift velocity.}\label{fig:SBMdrift}
\end{center}
\end{figure}

Figure \ref{fig:SBMdrift} shows the drift velocity streamlines in the case of the SBM.  Compared to Figure \ref{fig:PHMFdrift}, it is clear that the inward drift
from the poles is not as effective when employing the SBM as it was when using the Parker HMF:  the lines in the case of the SBM
fan out significantly more than in the case of the Parker HMF.  The value of $B_{\mathrm{T}}(b)/B_{\mathrm{R}}(b)=-0.1$, and has once again  been set to
a value low enough so that these differences can be clearly illustrated.

\subsection{The Effect on Modulation}

\begin{figure}[t!]
\begin{center}
\includegraphics[height=4.625 in,width=6.375 in, angle = 0]{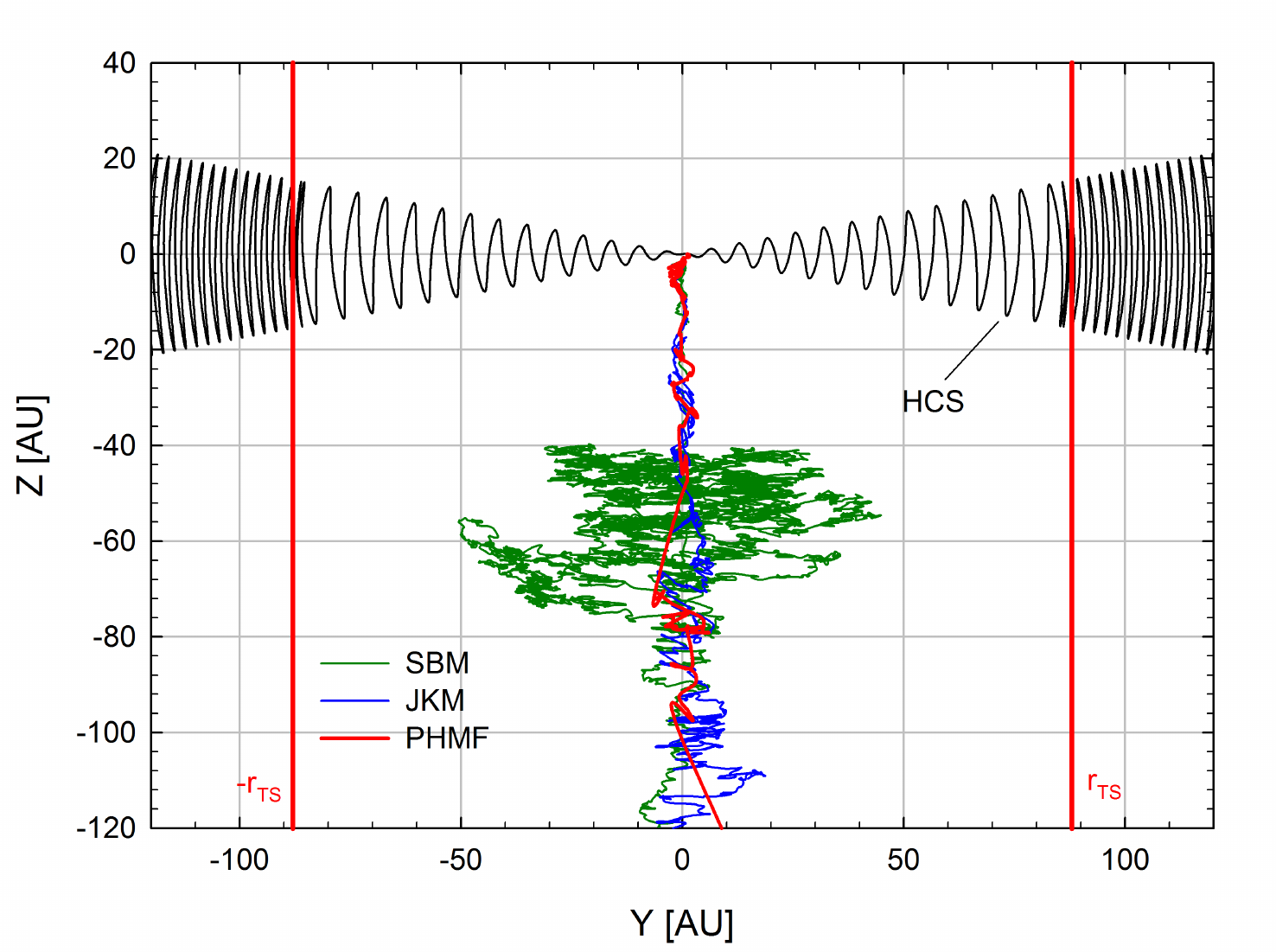}
\caption{\small \sl Pseudo-particle trajectories reflected in the meridional plane of the heliosphere for 1.5 GeV protons undergoing drift and diffusion during 
an $A>0$ drift cycle where $\alpha=10^\circ$.  The black line depicts the wavy HCS, while the red, green and blue lines respectively show the pseudo-particle trajectories 
through the polar regions
in each of the cases for the Parker HMF, the SBM, and the JKM.  The vertical red lines are drawn at the position of the TS, $r_{\mathrm{TS}}=88$ AU.
The Sun is located at $\left(Y,Z\right)=(0,0)$ with $Y$ the equatorial direction.}\label{fig:traj}
\end{center}
\end{figure}

Very effective use can be made of the properties of the SDE-based model to qualitatively illustrate the effect that these modifications to the HMF can have.  
Figure \ref{fig:traj} presents a meridional cut of the heliosphere and shows trajectories for pseudo-particles (1.5 GeV protons) entering in the polar regions 
during an $A>0$ polarity cycle, and propagating down towards the Sun, which is located at $\left(Y,Z\right)=\left(0,0\right)$;  the solid black line shows the 
warped HCS with tilt angle $\alpha=10^\circ$, for illustrative purposes.  The position of the TS, $r_{\mathrm{TS}}=88$ AU, is indicated by the vertical red 
lines.  The red, blue and green trajectories respectively shows the case for the Parker HMF, JKM and SBM, and the differences are significant:  in the case of the Parker HMF, 
the drift is obviously more effective, followed by consecutively reduced drift effects, i.e. reduced in terms of the ratio to diffusive processes, in the case of 
the JKM and SBM.  This is not only evident from the longer trajectories in the latter two cases, but also by the reduced occurrence of the long straight and 
jump-like segments in the trajectories, which indicate cases where $v_{\mathrm{d}}$ reaches unphysically large speeds.  From this figure, therefore, 
it is evident that both the JKM and SBM are effective in reducing drift effects over the poles and are thus successful in their original purpose.

\begin{figure}[t!]
\begin{center}
\includegraphics[height=4.625 in,width=6.375 in, angle = 0]{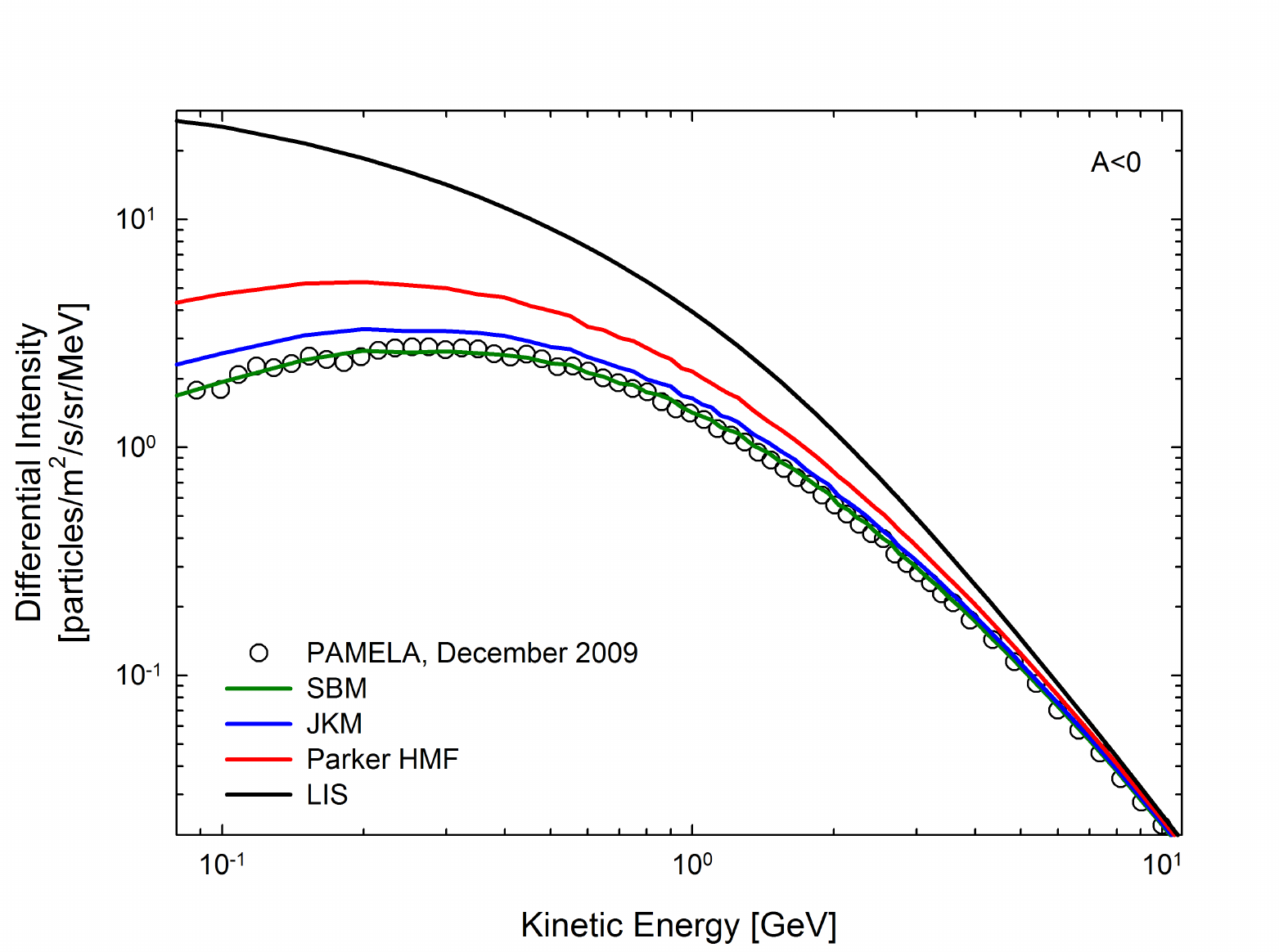}
\caption{\small \sl The SBM (green line) is applied to reproduce the 2009 Galactic proton spectrum as measured by PAMELA (circles) at Earth and published in e.g. Potgieter \textit{et al.} (2014).  
The blue and red lines respectively show the effects of the JKM and Parker HMF, using the same parameters as for the SBM.  The black line represents the LIS at $r=120$ AU.}\label{fig:PAM2009}
\end{center}
\end{figure}

Next, we look at the effect that these HMF modifications have on CR proton spectra;  in this way a quantitative measure of the effect on the modulation is obtained.
The model is used to reproduce the December 2009 CR proton 
spectrum, which was observed by the PAMELA experiment and averaged over 1 month (e.g. Adriani \textit{et al.}, 2013;  Potgieter \textit{et al.}, 2014).  This is 
shown in Figure \ref{fig:PAM2009}, where the observations from PAMELA are indicated by the circles and the model reproduction, assuming the SBM, 
is indicated by the green line;  the LIS at 120 AU is represented by the black line.  The values of all the relevant quantities for this model reproduction are indicated 
in the last column of Table~1.  To obtain a comparison with the SBM, these same values are used while assuming the Parker HMF (red line) and JKM (blue line) respectively;
the difference 
between these three HMF profiles is clearly depicted:  the JKM and SBM are consecutively lower than the Parker HMF;  these spectra peak in the vicinity
of 200 MeV and, at this energy, the SBM intensities are reduced relative to the LIS by about 86$\%$;  the JKM intensities are reduced by about 82$\%$ while, in the case of
the Parker HMF, intensities show a decrease of only about 71$\%$.
The fact that the spectrum resulting 
from the SBM is lower than that resulting from the JKM once again illustrates the fact that, although the effects of the JKM may be locally greater than that 
of the SBM, the overall effect of the SBM is larger than that of the JKM when integrated over the whole of modulation space.

\begin{figure}[t!]
\begin{center}
\includegraphics[height=4.625 in,width=6.375 in, angle = 0]{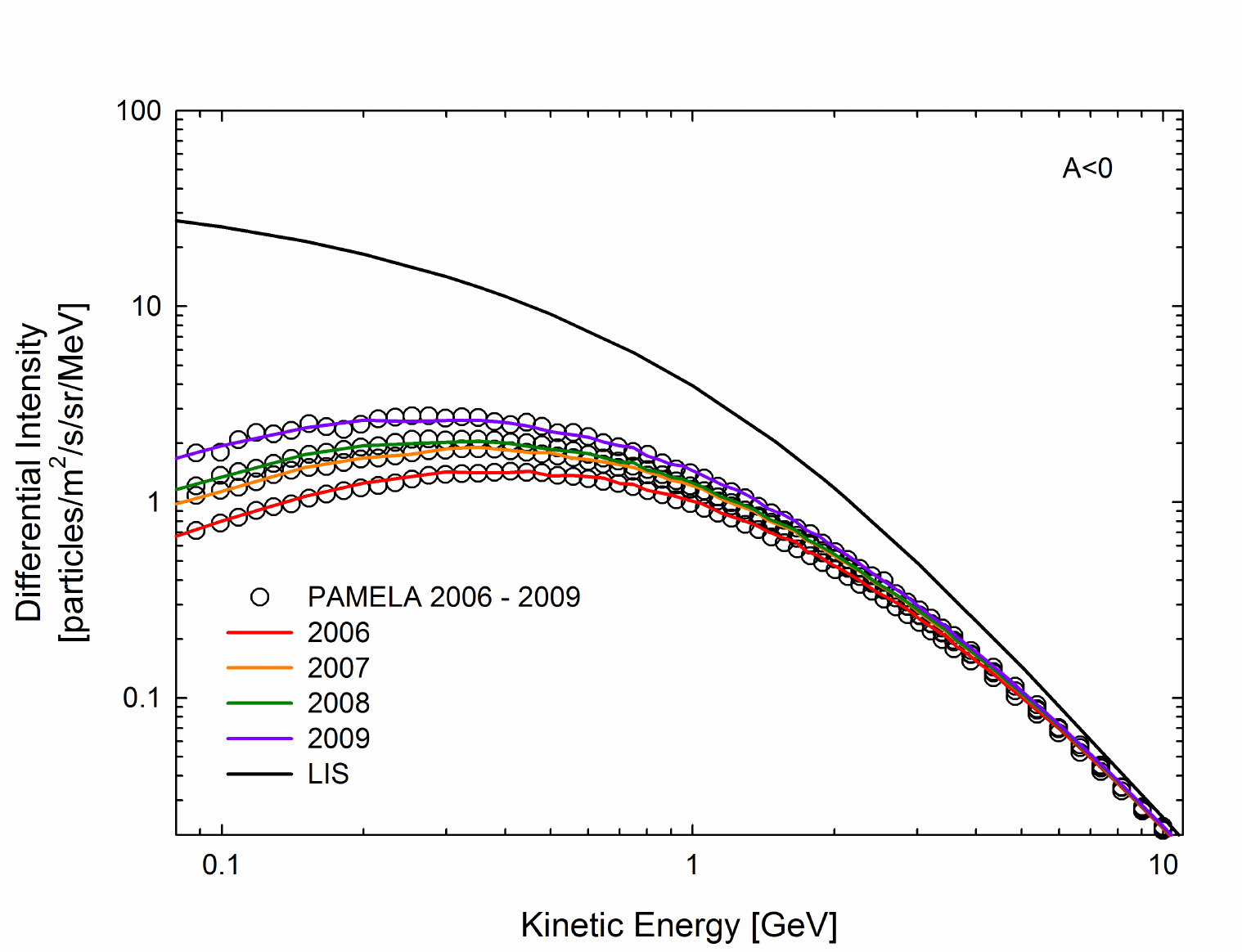}
\caption{\small \sl The circles represent the 2006 to 2009 Galactic proton spectra from observations, at Earth, by the PAMELA mission, as published by e.g. Potgieter \textit{et al.} (2014);
these spectra, from the lowest to the highest, correspond to November of 2006 and December of 2007, 2008, and 2009 respectively.  The SDE-based model of this study
is applied, utilising the SBM, to reproduce these observed spectra;  the reproductions for 2006, 2007, 2008, and 2009 are indicated by the red, orange, green and purple lines respectively,
and quite clearly become higher as well as softer toward 2009, with respect to the LIS (black line) at $r=120$ AU.}\label{fig:PAM2006-9}
\end{center}
\end{figure}

Further motivation for the applicability of the SBM comes from applying this modification to model the peculiar solar minimum stretching from 2006 to 2009.
Monthly averaged proton spectra measured by PAMELA at Earth, during November 2006 and December of 2007 and 2008 are now also reproduced using 
our model.  This is depicted in Figure \ref{fig:PAM2006-9}, the observations from PAMELA once again indicated by the circles, while the model reproductions
for 2006, 2007, 2008, and 2009 are indicated by the red, orange, green, and purple lines respectively;  the LIS at 120 AU is represented by the black line.  
The parameters to obtain these results are summarised in Table 1 and it 
is clear that:  1) the diffusion was required to continually increase from 2006 to 2009, with the value of $\kappa_{||,0}$ increasing from 16.0 in 2006 to 17.1, 
17.5, and 19.4 in 2007, 2008, and 2009 respectively;  2) the spectra became progressively softer towards 2009, so that the rigidity dependence of the diffusion 
coefficient below $\sim$ 3 GeV was required to change from $P^{0.85}$ in 2006 to $P^{0.81}$, $P^{0.78}$, and $P^{0.73}$ 2007, 2008, and 2009.  These 
type of changes are in agreement with those found by various other authors, including Bazilevskaya \textit{et al.} (2012), Ndiitwani \textit{et al.} (2013), Potgieter \textit{et al.} (2014), 
and Pacini and Usoskin (2015).  Except for providing evidence for the applicability of the SBM in particular, this qualitative correspondence in modulation results 
over the entire solar minimum period of 2006 to 2009 also illustrates the already well established credibility of the SDE-based model.

\begin{figure}
\begin{center}
\includegraphics[height=3.6 in,width=4.9 in, angle = 0]{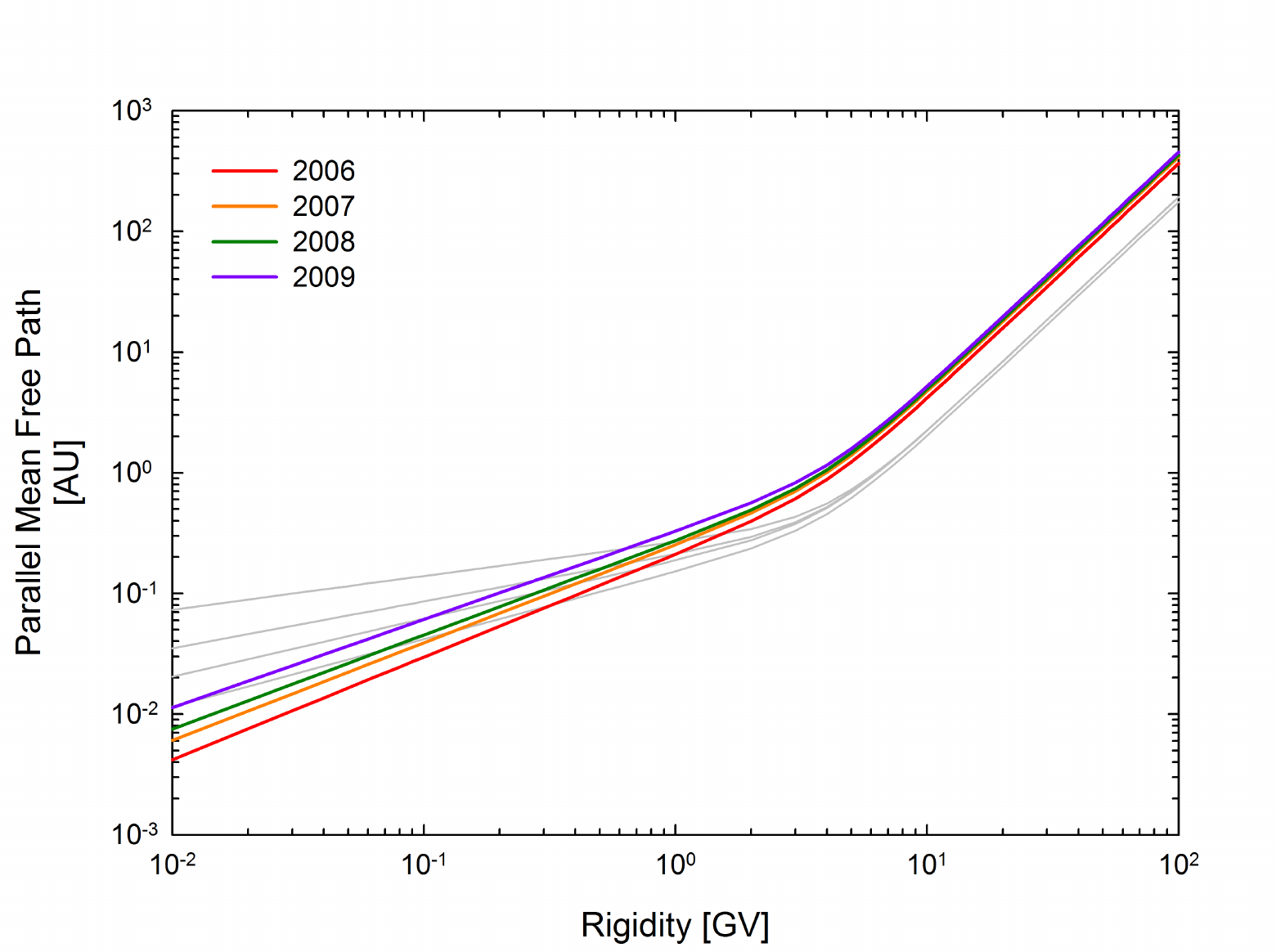}
\caption{\small \sl The parallel mean free paths $\lambda_{||}\mathrm{s}$ for 2006, 2007, 2008, and 2009 obtained from the SDE-based modeling shown in Figure \ref{fig:PAM2006-9} 
are represented by the red, orange, green, and purple lines respectively.  For comparison, the $\lambda_{||}\mathrm{s}$ that resulted from the modeling of 
Potgieter \textit{et al.} (2014) are included and represented by the grey lines, so that the qualitative correspondence between these two sets of results are clearly
depicted.}\label{fig:lambdaPAM}
\end{center}
\end{figure}

Figure \ref{fig:lambdaPAM} illustrates the results of Table 1 by depicting the mean free paths for each of 2006, 2007, 2008, and 2009.  The grey lines 
show the mean free paths obtained by Potgieter \textit{et al.} (2014) and, when compared to the results obtained in this study, the qualitative agreement is 
clearly illustrated.  Quantitative differences must exist between these two sets of results for three reasons.  Firstly, these two studies did not select
the same HMF profile, i.e. this study employs the SBM, while the study of Potgieter \textit{et al.} (2014) utilised the JKM;  referring to Figure \ref{fig:PAM2009},
this will clearly have a quantitative influence on the results.  Secondly, our model uses an updated LIS that was adapted according to observations by 
Voyager 1 after its crossing of the HP, as explained earlier.  Thirdly, there were some differences in the treatment of current sheet drift, the 
details of which are not relevant to the topic of this paper.

\begin{table}[]
\caption{Summary of the parameters used in this study to reproduce the 2006 to 2009 PAMELA proton spectra, using the SBM;  see
Eqs. (\ref{eq:PHMFspiral}), (\ref{eq:PHMF}) and (\ref{eq:SBM}).}
\centering
\begin{tabular}{l c c c c}
 \hline\hline
\textbf{Parameter} 										& \textbf{2006} 	& \textbf{2007} 	& \textbf{2008} 	& \textbf{2009} \\
\hline
$\alpha$ [deg]     										& 15.7          		& 14.0          		& 14.3          		& 10.0  	\\
$B_{\mathrm{e}}$ [nT]         								& 5.05          		& 4.50          		& 4.25          		& 3.94  	\\
$r_{\mathrm{TS}}$           								& 88.0	        	& 86.0	        	& 84.0          		& 80.0  	\\
\hline
$\kappa_{||,0}$ $[6\times 10^{20}$ cm$^2$.s$^{-1}]$		& 16.0			& 17.1	  		& 17.5			& 19.4 	\\
$a_1$													& 0.85			& 0.81			& 0.78			& 0.73  	\\
$P_0$ [GV]											& $1/\sqrt{10}$	& $1/\sqrt{10}$	& $1/\sqrt{10}$ 	& $1/\sqrt{40}$	\\
$P_{\mathrm{k}}$ [GV]									& 4.0			& 4.0			& 4.0			& 4.2		\\
\hline
\end{tabular}
\end{table}

\section{Conclusions}

An investigation into the effects of the SBM and JKM modifications were undertaken, making use of an SDE-based numerical modulation model.  These 
two modified HMF profiles were compared to that of the unmodified Parker HMF and it was illustrated that both these modifications change the radial 
dependence of the HMF in the polar regions, such as to bring about a larger HMF magnitude as function of radial distance.  This larger HMF magnitude 
in the polar regions then reduces both the diffusion and drift coefficients in these regions.  It was noted that, although the effects of the JKM may be 
locally larger than those of the SBM, the effects of the SBM are exercised over a greater part of modulation space and hence its overall effect on CR 
intensities is greater than that of the JKM.  The eventual effect on modulation of these two modifications were also illustrated and it was concluded that
both are effective in reducing drift effects over the poles;  this was illustrated by plotting drift velocity streamlines, as well as utilising the 
unique abilities of the SDE-based numerical model, which enables the construction of pseudo-particle trajectories.  It was then shown that the SBM can 
be used to reproduce the Galactic proton spectra observed by the PAMELA experiment during the peculiar solar minimum period of 2006 to 2009, and 
that the results thus obtained correspond qualitatively to that found by various other authors who have previously investigated this solar minimum.  This 
agreement was presented as veritable evidence for the applicability of the SBM in CR modulation studies;  it was also argued that the SBM is grounded 
in observational evidence and that, for these reasons, it should be preferred as modification to Parker type HMFs.

\section*{Acknowledgements}

MSP expresses their gratitude for the partial funding granted by the South African National Research Foundation (NRF) under the Incentive and Competitive Grants for 
Rated Researchers.  RDS thanks the NRF for financial support under Thuthuka Programme, grant number 87998.  JLR thanks the NRF and the South African Space Agency (SANSA) for partial financial 
support during his post-graduate study.  Opinions expressed and conclusions arrived at are those of the authors and are not necessarily to be attributed to the NRF.

\section{References}

\begin{description}
\item Adriani, O., \textit{et al.}:  2013, Astrophys. J. 765, 91 - 98.
\item Alanko-Huotari, K., Usoskin, I.G., Mursula, K., Kovaltsov, G.A.:  2007, J. Geophys. Res. 112, 1 - 10.
\item Bazilevskaya, G.A., Krainev, M.B., Makhmutov, V.S., Stozhkov, Yu.I., Svirzhevskaya, A.K., Svirzhevsky, N.S.:  2012, Adv. Space Res. 49, 784 - 790.
\item Bobik, P., \textit{et al.}:  2012, Astrophys. J. 754, 986 - 996.
\item Burger, R.A.:  2012, Astrophys. J. 760, 60 - 64.
\item Burger, R.A., Potgieter, M.S.:  1989, Astrophys. J. 339, 501 - 511.
\item Burger, R.A., Moraal, H., Webb, G.M.:  1985, Astrophys. Space Sci. 116, 107 - 129.
\item Burger, R.A., Potgieter, M.S., Heber, B.:  2000, J. Geophys. Res. 105, 27447 - 27456.
\item Dr\"{o}ge, W., Kartavykh, Y.Y., Klecker, B., Kovaltsov, G.A.:  2010, Astrophys. J. 709, 912 - 919.
\item Dunzlaff, P., Strauss, R.D., Potgieter, M.S.:  2015, Comp. Phys. Comm. 192, 156 - 165.
\item Effenberger, F., Fichtner, H., Scherer, K., Barra, S., Kleiman, J., Strauss, R.D.:  2012, Astrophys. J. 750, 108 - 115. 
\item Ferreira, S.E.S., Potgieter, M.S., Burger, R.A., Heber, B.:  2000, J. Geophys. Res. 105, 18305 - 18314.
\item Fichtner, H., le Roux, J.A., Mall, U., Rucinski, D.:  1996, Astron. Astrophys. 314, 650 - 662.
\item Fisk, L.A.:  1996, J. Geophys. Res. 101, 15547 - 15554.
\item Giacalone, J., Jokipii, J.R.:  1999, Astrophys. J. 520, 204 - 214.
\item Jokipii, J.R., Kopriva, D.A.:  1979, Astrophys. J. 234, 384 - 392.
\item Jokipii, J.R., K\'{o}ta, J.:  1989, Geophys. Res. Lett. 16, 1 - 4.
\item Kopp, A., B\"{u}sching, I., Strauss, R.D., Potgieter, M.S.:  2012, Comp. Phys. Comm. 183, 530 - 542.
\item K\'{o}ta, J., Jokipii, J.R.:  1983, Astrophys. J. 265, 573 - 581.
\item K\'{o}ta, J., Jokipii, J.R.:  1995, Int. Cosmic Ray Conf. Proc. 4, 680 - 683.
\item Langner, U.W.:  2004, Ph.D. thesis, Potchefstroom University for CHE, South Africa.
\item Langner, U.W., Potgieter, M.S.:  2004, J. Geophys. Res. 109, 1 - 12.
\item Luo, X., Zhang, M., Rassoul, H.K., Pogorelov, N.V.:  2011, Astrophys. J. 730, 13 - 22.
\item Luo, X., Zhang, M., Rassoul, H.K., Pogorelov, N.V., Heerikhuisen, J.:  2013a, Astrophys. J. 764, 85 - 100.
\item Luo, X., Zhang, M., Feng, X., Mendoza-Torres, J.E.:  2013b, J. Geophys. Res. 118, 7517 - 7524.
\item Ndiitwani, D.C., Potgieter, M.S., Manuel, R., Ferreira, S.E.S.:  2013, Int. Cosmic Ray Conf. Proc., icrc2013-0187.
\item Ngobeni, M.D., Potgieter, M.S.:  2008, Adv. Space Res. 41, 373 - 380.
\item Pacini, A.A., Usoskin, I.G.:  2015, Sol. Phys. 290, 943 - 950.
\item Parker, E.N.:  1958, Astrophys. J. 128, 664 - 676.
\item Parker, E.N.:  1965, Planet Space Sci. 13, 9 - 49.
\item Pei, C., Bieber, J.W., Burger, R.A., Clem, J.:  2010, J. Geophys. Res. 115, 1 - 12.
\item Phillips, J.L., \textit{et al.}:  1995, Geophys. Res. Lett. 22, 3301 - 3304.
\item Potgieter, M.S.:  2000, J. Geophys. Res. 105, 18295 - 18304.
\item Potgieter, M.S.:  2013, Living Rev. Solar Phys. 10, 3, 1 - 66.
\item Potgieter, M.S.:  2014, Braz. J. Phys. 44, 581 - 588.
\item Potgieter, M.S., le Roux, J.A., Burger, R.A.:  1989, J. Geophys. Res. 94, 2323 - 2332.
\item Potgieter, M.S., Haasbroek, L.J.:  1993, Int. Cosmic Ray Conf. Proc. 3, 457 - 460.
\item Potgieter, M.S., Nndanganeni, R.R., Vos, E.E., Boezio, M.:  2013, Int. Cosmic Ray Conf. Proc., icrc2013-0070.
\item Potgieter, M.S., Vos, E.E., Boezio, M., De Simone, N., Di Felice, V., Formato, V.:  2014, Sol. Phys. 289, 391 - 406.
\item Smith, C.W., Bieber, J.W.:  1991, Astrophys. J. 370, 435 - 441.
\item Sternal, O., Engelbrecht, N.E., Burger, R.A., Ferreira, S.E.S., Fichtner, H., Heber, B., Kopp, A., Potgieter, M.S., Scherer, K.:  2011, Astrophys. J. 741, 1 - 12.
\item Strauss, R.D., Potgieter, M.S., B\"{u}sching, I., Kopp, A.:  2011a, Astrophys. J. 735, 83 - 96.
\item Strauss, R.D., Potgieter, M.S., B\"{u}sching, I., Kopp, A.:  2011b, J. Geophys. Res. 116, 1 - 13.
\item Strauss, R.D., Potgieter, M.S., B\"{u}sching, I., Kopp, A.:  2012, Astrophys. Space Sci. 339, 223 - 236.
\item Yamada, Y., Yanagita, S., Yoshida, T.:  1998, Geophys. Res. Lett. 25, 2353 - 2356.
\item Zhang, M.:  1999, Astrophys. J. 513, 409 - 420.
\end{description}

\end{document}